\documentclass{article}
\usepackage{amsthm,xcolor,esvect,graphicx,verbatimbox,xspace,tikz,placeins,amssymb,hyperref,multirow,pgf-pie,pgfplots}
\usepackage[margin=1in]{geometry}

\newtheorem{definition}{Definition}

\newcommand{\D}{\mathcal{D}}
\newcommand{\E}{\mathcal{E}}
\newcommand{\disprop}{{\sf disprop}\xspace}

\newcommand{\seats}{S}
\newcommand{\votes}{V}
\definecolor{alizarin}{rgb}{0.82, 0.1, 0.26}
\definecolor{applegreen}{rgb}{0.55, 0.71, 0.0}
\definecolor{slategray}{rgb}{0.44, 0.5, 0.56}
\definecolor{amber}{rgb}{1.0, 0.75, 0.0}
\definecolor{mikadoyellow}{rgb}{1.0, 0.77, 0.05}
\definecolor{cadmiumgreen}{rgb}{0.0, 0.42, 0.24}
\definecolor{forestgreen}{rgb}{0.13, 0.55, 0.13}
\definecolor{lust}{rgb}{0.9, 0.13, 0.13}
\definecolor{denim}{rgb}{0.08, 0.38, 0.74}
\definecolor{purpleheart}{rgb}{0.41, 0.21, 0.61}
\definecolor{cherryblossompink}{rgb}{1.0, 0.72, 0.77}
\definecolor{darktangerine}{rgb}{1.0, 0.66, 0.07}
\definecolor{bananayellow}{rgb}{1.0, 0.88, 0.21}
\definecolor{lightblue}{rgb}{0.55,0.82,0.77}
\definecolor{lightgray}{rgb}{0.83, 0.83, 0.83}
\definecolor{languidlavender}{rgb}{0.84, 0.79, 0.87}
\definecolor{jasper}{rgb}{0.84, 0.23, 0.24}
\definecolor{teal}{rgb}{0.0, 0.5, 0.5}
\definecolor{anothergreen}{rgb}{0.4, 0.69, 0.2}
\definecolor{anotherblue}{rgb}{0.04, 0.73, 0.71}

\title{Redistricting for Proportionality}
\author{Moon Duchin, Gabe Schoenbach}
\date{December 2022}

\begin{document}

\maketitle


\begin{abstract}
American democracy is currently heavily reliant on plurality in single-member districts, or PSMD, as a system of election.  But public perceptions of fairness are often keyed to partisan  proportionality, or the degree of congruence between each party's share of the the vote and its share of representation.  PSMD has not tended to secure proportional outcomes historically, partially due to gerrymandering, where line-drawers intentionally extract more advantage for their side.  But it is now increasingly clear that even blind PSMD is frequently disproportional, and in unpredictable ways that depend on local political geography.  

In this paper we consider whether it is feasible to bring PSMD into alignment with a proportionality norm by targeting proportional outcomes in the design and selection of districts.  
We do this mainly through a close examination of the ``Freedom to Vote Test," a redistricting reform proposed in draft legislation in 2021. We find that applying the test with a proportionality target makes for sound policy: it performs well in legal battleground states and has a workable exception to handle edge cases where proportionality is out of reach.

\end{abstract}

\centerline{{\bf this is a preprint -- journal version now published at \href{https://www.degruyter.com/document/doi/10.1515/for-2022-2064/html?lang=en}{\tt {\color{blue} The Forum}}}}

\section{Introduction}

Most American legislative elections are plurality contests in single-member districts, a system we will abbreviate by PSMD:  the single candidate with the most votes is elected from each geographical district.  This stands in contrast to much of the rest of the world, where legislatures are filled by a mechanism that has proportionality guarantees by design, such as a ``party list" system that awards seats to political parties in proportion to their vote share.  The positive case for PSMD as a system centers on the importance of local and regional choice, with voters directly choosing their representatives; PSMD offers the prospect that preferences that are in the minority on a statewide level can secure majority support in a smaller geography.  However, it has long been observed that actual PSMD elections are routinely disproportional, and recent years have seen increasingly sophisticated modeling that shows that even blind PSMD---via randomized legally valid districts---is disproportional in ways that are unpredictable from state to state, because they depend on the precise spatial arrangement of vote preferences.  
Nevertheless, public perceptions of fair outcomes remain closely tied to  proportionality, i.e., the degree of congruence between vote share and seat share.  In this paper, we consider whether PSMD can be made compatible with a fairness norm calling for proportional outcomes.  
That is, it possible to counteract the central tendencies of party-blind districting and to expressly design districts that are likely to produce proportional representation?\footnote{In statistics, measures of {\em central tendency} include mean, median, and mode.  Variance and other measures of {\em dispersion} describe the concentration of a distribution, and  {\em the tails} is a broad term for the extreme values.}
Below, we address this question in highly practical terms, by considering how a test of gerrymandering could be administered in law, via the ``FTV Test" set down in draft Congressional legislation in 2021.  

That is, suppose a state were to adopt {\em partisan proportionality} as a goal in its law or guidelines for redistricting, reflecting a preference for a plan that is likely to elect a delegation with each party's seat share approximating its vote share.   Indeed, one state---Ohio---has proportionality written into its state constitution as a policy goal.\footnote{Note, however, that Ohio's redistricting following the 2020 Census has been a legal and procedural debacle, with this mandate completely ignored. The success of the reform measure at the ballot box at least gives us an instance of voter support for a proportionality ideal, though in this case ultimately disregarded, in defiance of the voters and the courts, by the commission composed of politicians.}  Is that a feasible goal for mapmakers?  Or is this an unattainable goal---and therefore an unreasonable legal standard?

After regaining control of both houses of Congress in 2021, Democratic lawmakers attempted several times to pass a bill that would address redistricting. In particular, they sought to rein in partisan gerrymandering in the wake of the 2019 {\em Rucho v. Common Cause} Supreme Court decision that declared it non-justiciable in federal courts.  Here is the proposed test from Senate Bill 2747, the ``Freedom to Vote: John R. Lewis Act" \cite{FTV}, paraphrased succinctly.
We will call it the {\bf FTV Test}.  (See Appendix~\ref{app:FTV} for full text.)

\begin{center}\begin{tikzpicture}

\draw [fill=alizarin, fill opacity=.65,rounded corners=8pt] (-8,-.75) rectangle (8,.75);
\node at (0,.225) {\em  According to a recognized metric of partisan fairness, a plan must yield nearly (within a set tolerance)} ;
\node at (0,-.225) {\em the {\bf ideal}/target seat share for each party in at least 3 out of a  prescribed list of 4 recent elections.};
\end{tikzpicture}\end{center}

The metric that sets the target was left unspecified, as we explore below, but 
the four elections were designated in the bill to be the two most recent contests for President and for U.S. Senate.
If the test is failed---i.e., if the tolerance is exceeded in two or more of the four contests---this triggers a {\em presumption} that the plan is gerrymandered, which can then be rebutted by the state.  
But as an initial matter, this is an unabashedly simple and results-based (not intent-based) test.  

In this paper we consider what kinds of fairness metrics are potentially suitable for use in the test, first discussing them in normative terms.  Circling back to the motivating question above, we focus on the viability of using the FTV Test with a proportionality target.
We focus on five states that had extensive legal challenges with a partisan gerrymandering component circa 2018:  North Carolina (challenged in federal and state court), Maryland (federal), Pennsylvania (state), Texas (federal), and Wisconsin (federal).\footnote{Texas was challenged for Voting Rights Act violations, i.e., on racial fairness grounds, but the complaints had a clear partisan element.}  We add Massachusetts, which is known to have a very uniform political geography, making the state a hard case for many fairness norms, as we will see below.
We are armed with large collections of alternative districting plans generated by computer as well as alternative plans drawn by courts and journalists, together with contemporaneous election data from the period preceding 2018. With this, we demonstrate that in the contested states, mapmakers can reasonably produce examples that satisfy the FTV Test for proportionality, suggesting that the test is not too stringent. 
Massachusetts give us an example where proportionality is out of reach, and is used here to illustrate that the FTV test handles that case well with the rebuttal element in its design.
But of course this proposed policy aims to validate plans for adoption that are likely to retain their good properties into the future.  This leads us to investigate whether plans that are near-proportional in an earlier election window will tend to remain so in subsequent years.  The findings there are encouraging as well.

\subsection*{Acknowledgments and disclosures}
The authors thank Eric McGhee for helpful exchanges about the efficiency gap standard and Max Fan and Chanel Richardson for invaluable data support.  We also thank Sarah Cannon, Devin Caughey, Daryl DeFord, Kenny Easwaran, Sam Hirsch, Michael Li, Ariel Procaccia, Jamie Tucker-Foltz, and Dave Wasserman for very useful feedback on this paper.  
MD served as an expert in {\em LWV v. Pennsylvania} in 2018; this role was one of analysis and evaluation, and MD did not design any plan presented below.

\section{Proportionality and other ideals}

\subsection{Norms of fairness}

By design, the FTV Test leaves open an obvious question---what is the ideal seat share?  
Several possibilities might be considered.  Let us denote by $V$ the statewide share of the (major-party) vote that belongs to a political party, and by $S$ their share of representation, for a given vote pattern and districting plan.

\newpage
\paragraph{Normative possibilities for representational outcomes}
\begin{itemize}
\item {\bf Proportionality.}  The ideal seats outcome in a given election is $S=V$.  

{\em Normative rationale: democracy means that public preferences should be converted to representation.  This standard keys representation directly to the share of votes secured. As a standard for single-member districts, it seeks for the delegation as a whole to reflect statewide preferences while individual representatives reflect local preferences.}

\item {\bf Efficiency gap.}\footnote{To be precise, this is the simplified efficiency gap advanced by Eric McGhee in \cite{SEG} (and in personal communication) as a viable alternative to the original score.  The original efficiency gap defined in \cite{OEG} is almost the same, but with what amounts to a noise term that depends on the turnout disparity across the districts, as McGhee notes.  See \cite{Veomett} for an extended analysis of the turnout noise.}      The ideal seats outcome in a given election is $S=2V-\frac 12$.

{\em Normative rationale: some amount of bonus for the party with more support may provide a more effective ability to legislate, particularly to avoid near-parity and a corresponding deadlock.
In addition, an amplified advantage for the party with more support can make a system more responsive to shifts in voter preference, which inclines against entrenchment of incumbents.
This standard proposes a {\em double}-bonus, in which every additional point of vote share deserves {\em two} points of seat share.  This slope of two also corresponds to equalizing the parties' ``wasted votes" under some simplifying assumptions.}
\item {\bf Ensemble mean.}  The ideal seats outcome in a given election is the average $S$ over all the districting plans in an algorithmically generated ``ensemble" of comparator plans.

{\em Normative rationale: we are committed to districts because they can secure regional or neighborhood-specific representation, because they may allow for minority representation within a plurality system, and/or because they are supported by a long history.  Since we want to eliminate distortions caused by partisan control of the lines, we should key our notion of fairness to the typical  consequences of party-blind redistricting, effectively converting a procedural norm to a substantive norm.}
\end{itemize}

We note that there are quite a few other notions of partisan fairness in the literature, in addition to other ways to operationalize the norms cited here.\footnote{For instance, we could adopt an alternative arithmetic of wasted votes to obtain any slope $S=mV-\frac{m-1}2$ for the efficiency gap; see \cite{Cover}. And some authors in the game theory and fair division literature have argued for a different target derived from ensembles, such as averaging the ensemble min and max rather than using the mean. (This is explicitly argued in \cite{BPTF}, crediting \cite{LandauSu}, where it is called a ``geometric target" because it takes spatial arrangements---which we would term political geography---into account.)} The {\em partisan symmetry} family of metrics (which includes the mean-median gap, the partisan bias score, and so on) is particularly prominent, but these metrics do not identify a target or ideal seat share in a given election.  Advocates of partisan symmetry, principally Gary King and collaborators, argue that this feature---that the scores are not prescriptive of seat outcomes---is a conceptual strength (see \cite{KKR} and its references). Whether or not this is so, it means that symmetry scores can not be plugged into an outcome-oriented evaluation test like this one.\footnote{Furthermore, a recent paper of the current authors and several collaborators \cite{PSymm} shows that symmetry scores are typically correlated poorly if at all with outcome-oriented metrics like the ones considered here.  Indeed, symmetry scores are prone to numerous problems and paradoxes from the point of view of seats outcomes.}   The same holds for {\em declination} \cite{War}, which is denominated in an abstract trigonometric unit (the arctangent of a certain angle in a plot of votes by district) that has no direct relationship to the seats outcome.\footnote{And there are many more, with some that also take geography into account.  These include DeFord--Eubank--Rodden's {\em  partisan dislocation}, which compares a voter's district to that voter's nearest neighbors \cite{AAPD}; Eguia's {\em artificial partisan advantage}, which compares a voter's district to their county \cite{Eguia}; and Campisi--Ratliff--Somersille--Veomett's {\em GEO} metric \cite{GEO}, which compares each district to its neighboring districts.  An ideal seat share could be derived from partisan dislocation and artificial partisan advantage, but we do not attempt that here.}

Reasonable people could differ about which normative argument listed above is the most persuasive.  We argue here for proportionality. The definition is straightforward, the intuitive case for the norm is clear, and there are no arbitrary or black-box elements in the construction of the standard.  

By contrast, the slope of two in the efficiency gap standard is arguably somewhat arbitrary and in any case has counter-intuitive effects:  for example, a proportional outcome is often labeled as a gerrymander according to $EG$ (for instance, if a party has 60\% of the seats and 60\% of the representation, i.e.,  $S=V=.6$, this registers as a significant gerrymander {\em against} that party).  Secondly, if a state had a 75-25 voting advantage for one party, then no matter how large the delegation, efficiency gap advocates would find a 100\%-0\% sweep of the seats to be ideal.  (For a discussion of general features of the $EG$ metric, see for instance \cite{EG-Notices}.)

There are at least two obstructions to enshrining an ensemble-based standard, like the ensemble mean described above, in a simple litmus test for gerrymandering.  One is that there are many details of ensemble construction that can have subtle impacts on the representational outcomes.  (Should county preservation be encouraged in the algorithm that draws districts, or not?---and if so, how?  If the modeler aims to draw a representative sample of plans, from what probability distribution should it be drawn?  and so on.)  As ensemble practitioners  ourselves, we find the method to be best suited to holistic hypothesis testing about the consequences of different operational frameworks of rules; ensembles are far less suitable for deriving a manageable (still less a canonical) ideal outcome.  Without a close look at the design choices of the modeler, ensembles risk becoming a high-stakes black box.  The second major issue is the glaring question of whether it is normatively sound to promote the central tendencies of PSMD to the status of an ideal.  Districting algorithms can produce plans that are {\em neutral} with respect to partisan (and racial, and other) data, but clearly the facial neutrality of a procedure is no guarantee of the fairness of the outcome in any larger social or political sense.\footnote{In case it seems counter-intuitive that neutral processes may not give fair outcomes, consider a scenario in which a limited amount of food must be portioned out to two people:  a small child and a large adult.  It is procedurally neutral to give them equal portions, or random portions, but it seems clearly preferable to pursue the overriding norm that each one should receive adequate nutrients and neither one should go hungry (insofar as this is possible given the available resources).}
Indeed, if we are persuaded from first principles that proportionality is a healthy goal, then well-designed ensembles give us a measure of just how much blind-PSMD falls short of fairness.
In this article, we will use the method of ensembles to generate a batch of plans that we can scan for some desirable properties---that is, the algorithms are mainly engines of examples---but we will studiously resist the move that declares the typical outcome to be the ideal.

\subsection{Quantitative formulation of disproportionality and near-proportionality}
\begin{definition}
    Given a districting plan $\D$ and an election $\E$, the \textbf{disproportionality} (from the Republican point of view, say) is the difference between the seat share and the vote share.  Adopting notation to remind ourselves that these quantities depend both on $\D$ and on $\E$, we can write
    $$\disprop=\seats-\votes, \qquad \hbox{or more fully,} \qquad  \disprop(\D, \E) = \seats(\D,\E) - \votes(\E),$$ 
where $\seats=\seats(\D,\E)$ is the share of districts that have Republican majorities in $\D$ under $\E$ and $\votes=\votes(\E)$ is the statewide Republican vote share.  This score obtains its ideal 
when $\disprop=0$, i.e., when $\seats=\votes$.  
\end{definition}

For example, in North Carolina ($k = 13$ before the 2020 reapportionment), the 2016 Presidential election had $\votes \approx 0.520$, i.e., very nearly 52\% Republican vote share statewide. Under that cast vote pattern, 10/13 of the Congressional districts had Republican majorities, so the enacted plan under Pres16 has a disproportionality  of $10/13 - 0.520 = 0.249$, meaning that Republicans got more seats than their vote share by a huge margin, amounting to nearly a quarter of the delegation. This is a signed metric; a negative \disprop would similarly indicate Republican disadvantage. While this score is built from just one election pattern, we will make use of as many statewide contests as are available---serially, not by averaging---to best understand the political geography of the state. Table~\ref{tab:NC-enacted} shows statistics from nine elections from the period leading up to 2018.

\begin{table}[ht]
\begin{tabular}{cc|ccccccccc}
&election & Gov08 & Sen08 & Sen10 & Gov12 &
Pres12 & Sen14 & Pres16 & Sen16 & Gov16 \\
&R vote share & .483 & .457 & .560 & .559 &
.511 & .508 & .520 & .530 & .500 \\
\hline \hline 
\multirow{3}{*}{\rotatebox{90}{{\em targets}}}&proportional&$6.3$ & $5.9$ & $7.3$ & $7.3$ &
$6.6$ & $6.6$ & $6.8$ & $6.9$ & $6.5$ \\
\cline{2-11}
&$EG=0$&$6.0$ & $5.4$ & $8.1$ & $8.0$ &
$6.8$ & $6.7$ & $7.0$ & $7.3$ & $6.5$ \\
\cline{2-11}
&ensemble mean& 4.3 & 3.0 & 10.3 & 9.7 & 7.3 & 7.6 & 7.8 & 8.5 & 7.3\\
\hline
\hline
\multirow{3}{*}{\rotatebox{90}{Leg12}}& R seats & 8 & 10 & 10 & 10 & 10 & 10& 10& 10 & 10 \\
\cline{2-11}
&R seat share & $.615$ & $.769$ & $.769$ & $.769$ &
$.769$ & $.769$ & $.769$  & $.769$ & $.769$  \\
\cline{2-11}
&\textbf{\disprop} & $.132$ & $.312$ & $.209$ & $.211$ &
$.258$ & $.261$ & $.249$ & $.239$ & $.270$ \\
\hline
\hline
\multirow{3}{*}{\rotatebox{90}{Leg16}}& R seats & 7 & 5 & 10 & 10 & 10 & 10& 10& 10 & 10 \\
\cline{2-11}
&R seat share & $.538$ & $.385$ & $.769$ & $.769$ &
$.769$ & $.769$ & $.769$  & $.769$ & $.769$  \\
\cline{2-11}
&\textbf{\disprop} & $.055$ & $-.072$ & $.209$ & $.211$ &
$.258$ & $.261$ & $.249$ & $.239$ & $.270$ \\
\hline
\hline
\multirow{3}{*}{\rotatebox{90}{Judges}}& R seats & 4 & 4 & 9 & 9 & 8 & 9& 8& 9 & 8 \\
\cline{2-11}
&R seat share & $.308$ & $.308$ & $.692$ & $.692$ &
$.615$ & $.692$ & $.615$  & $.692$ & $.615$  \\
\cline{2-11}
&\textbf{\disprop} & $-.175$ & $-.149$ & $.132$ & $.134$ &
$.105$ & $.184$ & $.096$ & $.162$ & $.116$ \\
\hline
\hline
\multirow{3}{*}{\rotatebox{90}{538Pro}}& R seats & 5&	6&	9&	7&	7&	7&	7&	7&	7 \\
\cline{2-11}
&R seat share& .385&	.462&	.692&	.538&	.538&	.538&	.538&	.538&	.538 \\
\cline{2-11}
&\textbf{\disprop} & $-.098$	&$.005$	&$.132$	&$-.020$	&$.028$	&$.030$	&.019	&.008	&.039 \\
\hline
\end{tabular}
\caption{Election results from a series of  ``up-ballot" statewide contests $\E_1,\dots,\E_9$ in North Carolina (with $k=13$ seats), laid over four districting plans: the legislature's enacted plan from 2012, their proposed remedial plan from 2016, a plan designed by a bipartisan panel of retired judges without partisan data, and a plan designed for a journalistic project with the goal of proportionality. The table shows the ideal seats target out of 13 for each standard.  Comparing targets under Sen10 ($V=.560$) and Gov12 ($V=.559$) reflects the fact that the ideal specified by ensemble mean is sensitive to the specific geography of the votes, while the other two targets depend only on $V$. 
Comparing Sen08 and Sen10 shows that the $EG$ standard is more responsive to vote swings than proportionality, while ensembles in this case are arguably over-responsive.
``R seats" counts districts in each plan with more R votes than D votes in the given election pattern.}\label{tab:NC-enacted}
\end{table}

\begin{figure}[htb!]
    \centering
\begin{tikzpicture}

\node at (0,0) {\includegraphics[width=3.8in]{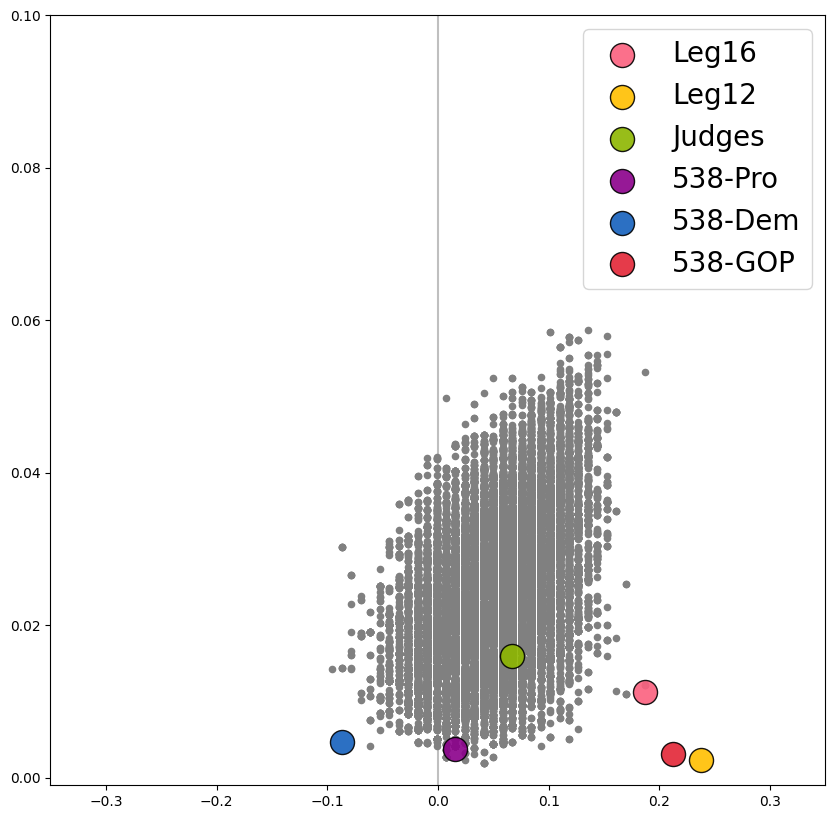}};
\node at (0,-5.3) {Mean \disprop};
\node at (-5,-5.1) {\small $\longleftarrow$ D-favoring};
\node at (5,-5.1) {\small  R-favoring $\longrightarrow$};
\node at (-5.5,0) [rotate=90] {Variance};
\end{tikzpicture}    
    \caption{Statistics on disproportionality from North Carolina, with a party-neutral ensemble of 100,000 alternative plans plotted in gray. Handmade plans are marked with colored dots. Note that both legislative plans have high disproportionality, while the Judges' plan is much closer to a typical map in the ensemble. The 538 Proportional plan succeeds at being very proportional with low variance, even though it was made using a different vote index.  (This figure is repeated in Figure~\ref{fig:all-scatters} below, alongside corresponding plots for the five other states studied here.)}
    \label{fig:NC-scatter}
\end{figure}
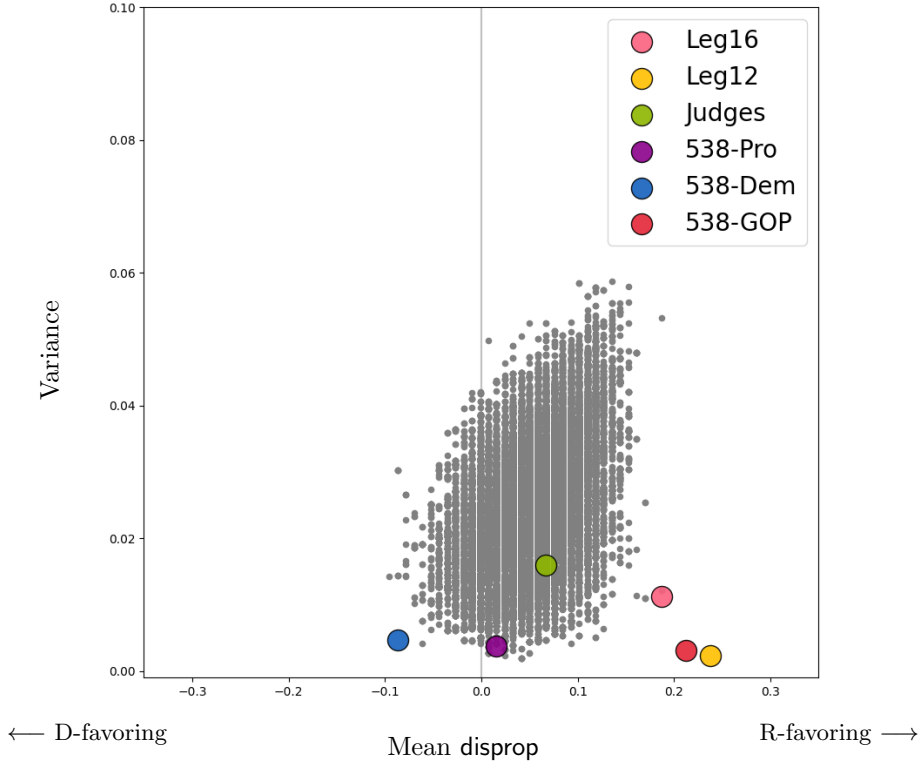

Given our districting plan $\D$, studying disproportionality longitudinally, over the course of a suite of elections,  can help to highlight trends. 
In particular, if the \disprop values have high variance, we can regard the plan as being quite volatile with respect to proportionality, such as by favoring one party in some elections and the opposite party in others.
If the values have high magnitude and low variance (which implies a consistent sign), then the plan strongly and consistently favors one party.
In order to understand whether the means and variances are high or low in context of reasonable North Carolina alternatives, we then show in Figure~\ref{fig:NC-scatter} how they compare to an ensemble of 100,000 alternative plans created by a race- and party-neutral process.\footnote{See Appendix~\ref{app:recom} for a description of the {\em recombination} procedure used to draw these districts.}  
We include in the comparison a variety of handmade plans as points of comparison, including plans made by journalist and political analyst Dave Wasserman as part of the 538 Atlas of Redistricting project \cite{538}.  Wasserman designed plans with the intention of demonstrating Democratic gerrymanders (538-Dem), Republican gerrymanders (538-GOP), and proportionality-favoring (538-Pro) alternatives in each state.  The 538 plans help to illustrate that a conventional mapmaker who seeks to promote some objective, drawing by hand and without algorithmic assistance, can often rival or surpass the outputs of a large neutral ensemble generated by computer.\footnote{There is no contradiction here, since ensembles should be thought of as illustrating the  central tendencies of a given {\em redistricting problem} (districts with specified sizes and properties interacting with given geography and vote patterns).  They often find configurations that people do not, but they are best for exploring the main bulk of likely outcomes; by contrast, related optimization techniques can be used to explore the {\em tails} or extremes.  See, for instance, \cite{shortbursts}.}
This should raise our confidence that if (say) ten or twenty percent of a neutral ensemble has a desirable property, it will correspond to a feasible ask for a skilled mapmaker, even without hyper-partisan tuning.

Disproportionality scores themselves do not vary continuously, but by jumps, since the $\seats$ term must be one of $0,\frac 1k, \frac 2k, \dots, 1$ for a delegation of size $k$. In particular, it will typically not be possible to get the seat share to exactly match the vote share.  So instead of demanding a perfect score, we should hope that a plan scores a disproportionality sufficiently close to zero on a sequence of past elections.

\begin{definition}[Near proportionality]
Given a threshold $t > 0$, a plan-election pair $(\D, \E)$ is \textbf{near-proportional} if $|\disprop(\D, \E)| < t$. 
\end{definition}

This definition will help us study the FTV Test, which employed a threshold for Congressional plans of whichever is greater, one seat or 7\%.  
That is, the language of the bill specified $t = \mathsf{max}(0.07, \frac 1k)$, where $k$ is the number of seats. A proposed districting plan would be flagged as ``materially favoring" a political party if its disproportionality exceeds $t$ for at least two out of four recent Presidential and Senate elections. In other words, a plan needs to pass for at least three out of these four contests to pass the test overall.   For example, if North Carolina's enacted plan had been assessed in 2018 as part of the litigation that year, its pass rate in the previous two Senate and two Presidential races would be zero out of four, since the pro-Republican disproportionality is over 23\% in all four elections, while the allowed threshold would be only 7.7\% (one seat out of 13).  
The Judges' plan reduces the disproportionality in every one of these four elections, but nevertheless also misses the mark all four times.  This may not be surprising in light of the fact that the Judges did not have a mandate to design for proportionality, but instead to largely disregard partisanship.  The ensemble comparison in Figure~\ref{fig:NC-scatter} gives us evidence that they succeeded.

In this paper we focus on assessing the soundness of the FTV Test, granting its choice of elections $\E_1,\dots,\E_4$ and threshold $t$ and considering possible standards to set the target for representation.  We attempt to show that it is not too stringent and not too loose, applying a kind of {\em Goldilocks principle}.\footnote{See, for instance, \cite{Compet} and \cite{BPTF}, where authors consider other gerrymandering tests in Goldilocks terms.}  For us, the evidence that supports a positive Goldilocks conclusion is that at least 10\% of a blind ensemble passes and at least 10\% of a blind ensemble fails in every state but Massachusetts---and the test has an out that would work well in that hard case. 
Furthermore, the FTV Test and the use of a neutral ensemble of computer plans correctly handles the plans that came to us with a ground-truth label as gerrymandered, party-blind, or proportionality-seeking.  
It would be interesting to also consider how the choice of elections and thresholds interacts with the standard to make a sound test, but we leave further considerations in this direction to future work.\footnote{For instance, $S/V$ or $(S-V)^2$ could be used to measure deviation from proportionality, though these sacrifice some of the interpretability of the simple difference $S-V$.
And the threshold $t=\max(.07,1/k)$ could certainly be tweaked, such as by using $1/2k$ instead of $1/k$ to demand outcomes that are as close as possible to the target rather than allowing one seat of latitude in either direction.  The value $.07$ itself is certainly arbitrary, but recalls the allowed deviation of $.07$ or $.08$ that has sometimes been proposed for use with the efficiency gap \cite{OEG}.  The effect here is to allow more latitude than one seat in states with at least $15$ districts, which is the level at which $1/k<.07$.  From the 2010 to the 2020 Census, this held steady at seven states: CA, TX, FL, NY, IL, PA, and OH.}  

\section{Validating proportionality on past elections}

Two questions are immediately posed by this discussion:  first, how easy is it to design a near-proportional plan, even with perfect knowledge of the set of elections to be used in the test?  And second, is past proportionality likely to predict future proportionality?

\subsection{How attainable is near-proportionality?}
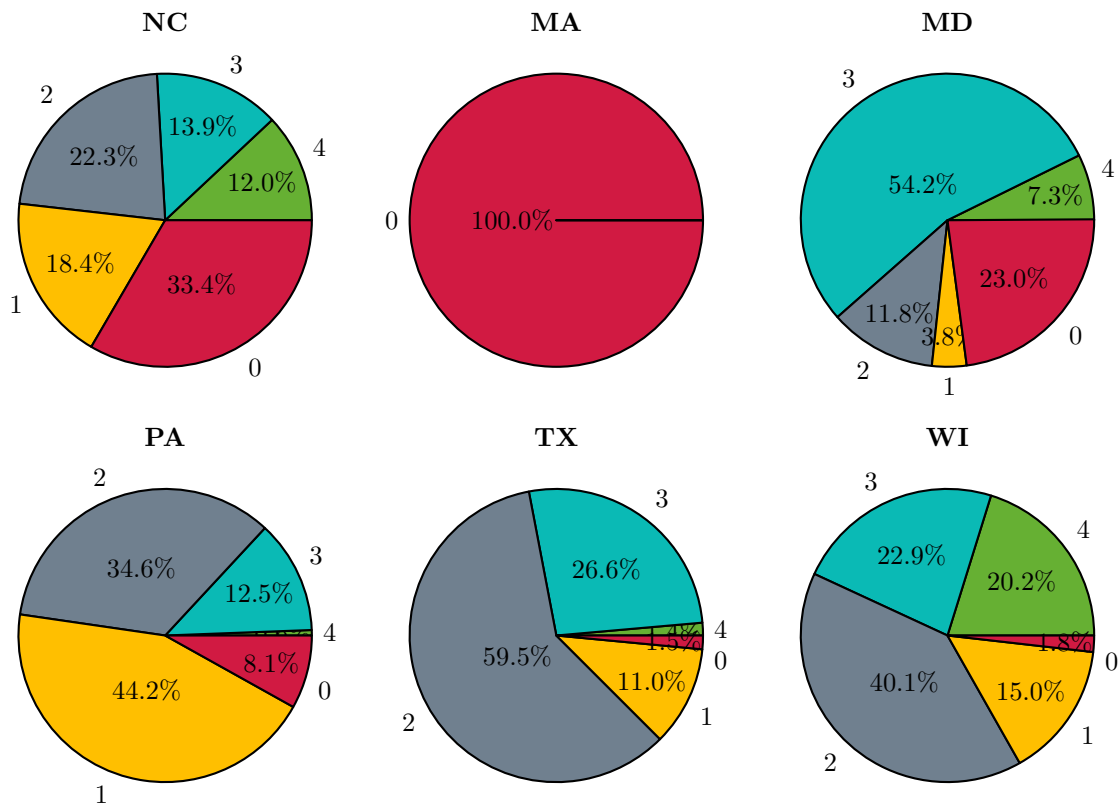
\begin{figure}[htb!]
    \centering

\begin{tikzpicture}[scale=.65]
\pie[pos={0,8.5},color={anothergreen,anotherblue,slategray,amber,alizarin}]{12.0/4,13.9/3,22.3/2,18.4/1,33.4/0} 
\node at (0,8.5) [above=2.4cm] {\bf NC};
\pie[pos={8,8.5},color={alizarin}]{100.0/0}
\node at (8,8.5) [above=2.4cm] {\bf MA};
\pie[pos={16,8.5},color={anothergreen,anotherblue,slategray,amber,alizarin}]{7.3/4,54.2/3,11.8/2,3.8/1,23.0/0}
\node at (16,8.5) [above=2.4cm] {\bf MD};
\pie[pos={0,0},color={anothergreen,anotherblue,slategray,amber,alizarin}]
{0.6/4,12.5/3,34.6/2,44.2/1,8.1/0}
\node at (0,0) [above=2.4cm] {\bf PA};
\pie[pos={8,0},color={anothergreen,anotherblue,slategray,amber,alizarin}]
{1.4/4,26.6/3,59.5/2,11.0/1,1.5/0}
\node at (8,0) [above=2.4cm] {\bf TX};
\pie[pos={16,0},color={anothergreen,anotherblue,slategray,amber,alizarin}]
{20.2/4,22.9/3,40.1/2,15.0/1,1.8/0}
\node at (16,0) [above=2.4cm] {\bf WI};
\end{tikzpicture}

    \caption{Breakdown of 100,000 districting plans on each of NC, PA, WI, MA, MD, and TX, scored by near-proportionality using the most recent two Presidential and Senate races with $t=\max(.07,1/k)$, as in the FTV Test. Each slice of the pie shows how many plans had a near proportionality score of X/4, with X ranging from 0 (\textbf{\color{alizarin}{red}}) to 4 (\textbf{\color{anothergreen}{green}}), counter-clockwise around the circle.  For each state, the \textbf{\color{anotherblue}{blue}} and \textbf{\color{anothergreen}{green}} sectors show the share of plans that would pass the test with three or four successes, respectively.  More than a quarter of neutrally-drawn plans pass the test in North Carolina, Maryland, Texas, and Wisconsin; in Maryland, it's well over half. Thousands pass in Pennsylvania, while none at all do so in Massachusetts.}
    \label{fig:pie}
\end{figure}

Figure \ref{fig:pie} shows 100,000 population-balanced plans in each of six states, broken down by their success rate for near-proportionality in the four indicated elections.

In the states from our dataset, we can see appreciable differences in whether the central tendencies of PSMD support the selection of plans with near-proportional historical outcomes.  
It is fairly common (25-30\% frequency) for our process to generate a near-proportional plan by chance in North Carolina and Texas, while enforcing only contiguity, population balance, and a preference for compactness.  
In Maryland and Wisconsin, it is even more frequent, occurring over 40\% of the time in our sample.    
The  Pennsylvania sample gives over 13,000 examples of compliant plans, i.e., more than one in eight plans that were generated passes the test.  
(Recall that only a {\em single} compliant plan is ultimately needed, with no badge of typicality required in the test.)  Finally, in Massachusetts, a proportional plan simply never occurred in our sample.\footnote{This is expected behavior in Massachusetts.  Duchin et al. showed in \cite{Mass-ELJ} that, even though Republican votes have frequently occurred in significant numbers, they are so uniformly distributed around the state that Republicans will typically have major representational shortfalls, and will sometimes be entirely locked out of the possibility of representation.}

Accordingly, ensemble evidence could clearly be used to protect an FTV-failing plan in Massachusetts from invalidation.
Pennsylvania might attempt to rebut the presumption of a gerrymander, arguing that proportionality is difficult to secure.  But on this reading of the bill's intent, that attempt would fail and the other four states would have still more difficulty defending failing plans.

\FloatBarrier

\begin{figure}[ht]
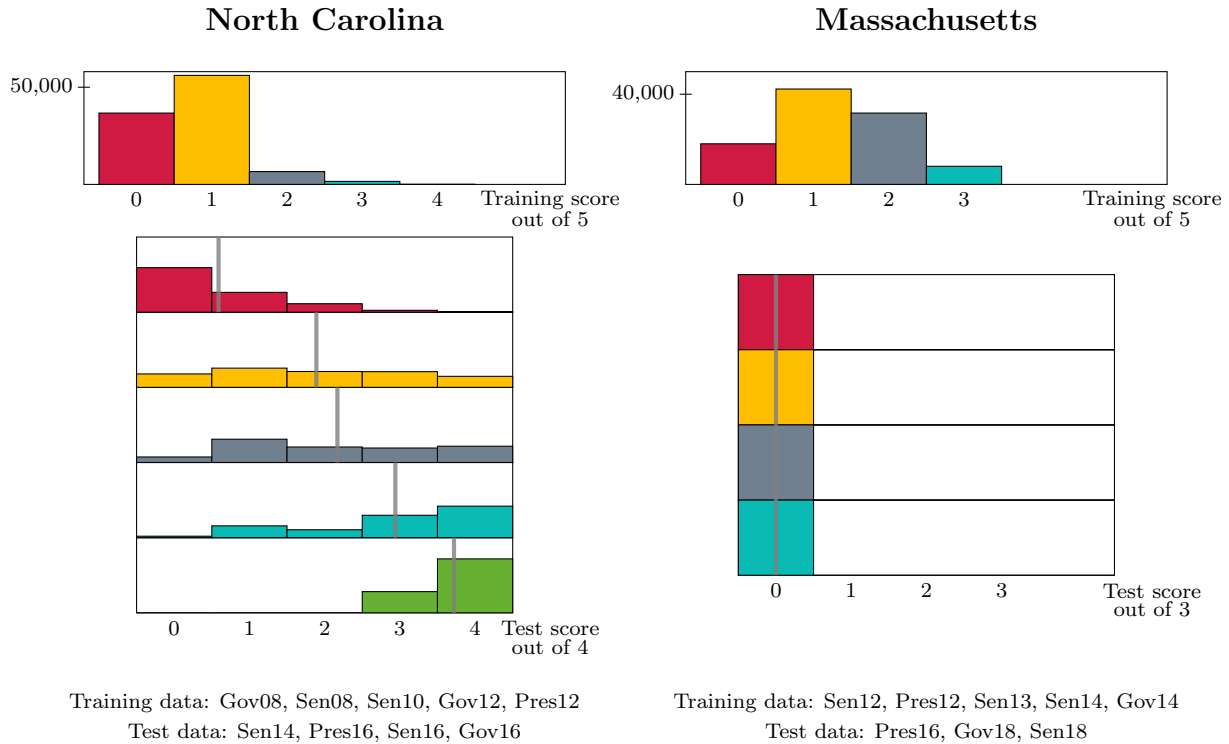
\centering
\begin{tikzpicture}
\begin{scope}[xshift=0,yshift=11cm]
\input{NC-stack}
\end{scope}
\begin{scope}[xshift=8cm,yshift=11cm]
\input{MA-stack}
\end{scope}
\end{tikzpicture}
\caption{A simple visualization of the predictive power of near-proportionality.  Here, the training score is the number of near-proportional elections at $t=.07$ from an earlier set of contests.  The test score shows the number of near-proportional elections with the same threshold for a later set of elections.  In North Carolina, the proportionality performance on the later elections improves steadily as the earlier performance is increased.  However, Massachusetts is different, as usual.  The 2012--2014 period includes some election geography that makes it possible to achieve proportionality, so some maps earn a good training score.  However, the 2016 and 2018 elections revert to form for the state, affording no instances of near-proportionality.}\label{fig:hists1}
\end{figure}

\subsection{How predictive is past proportionality?}
\begin{figure}
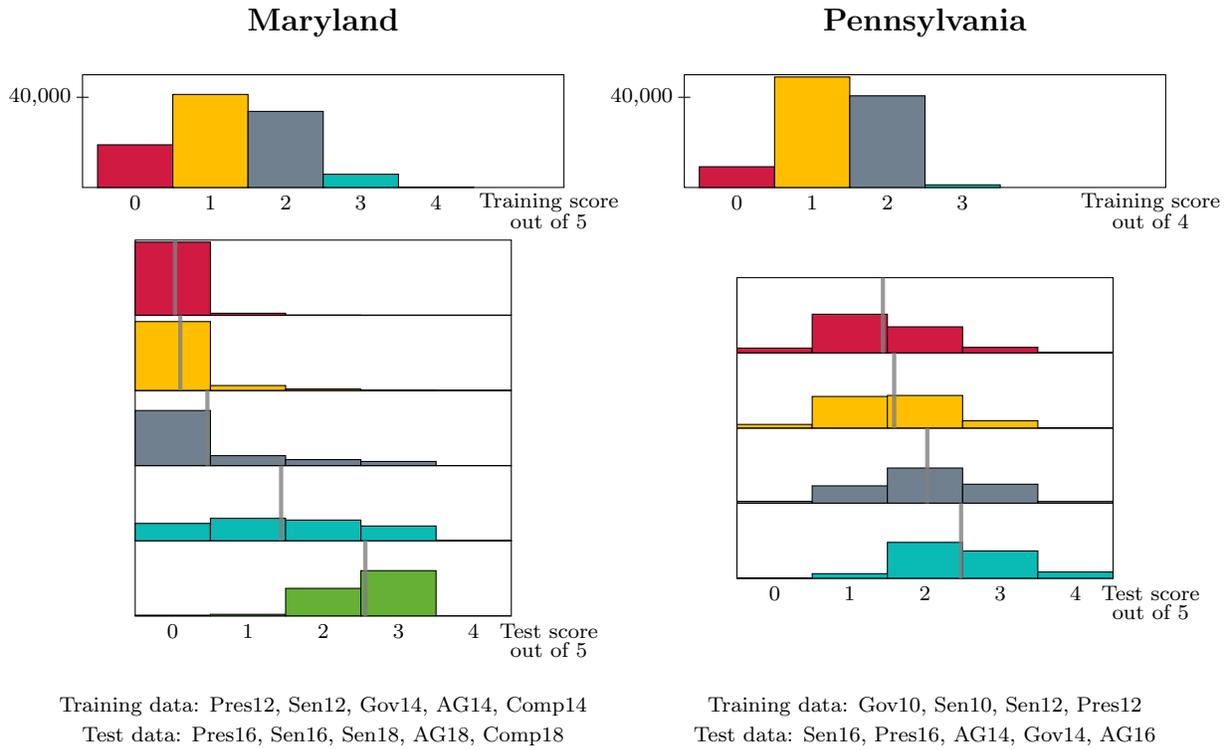
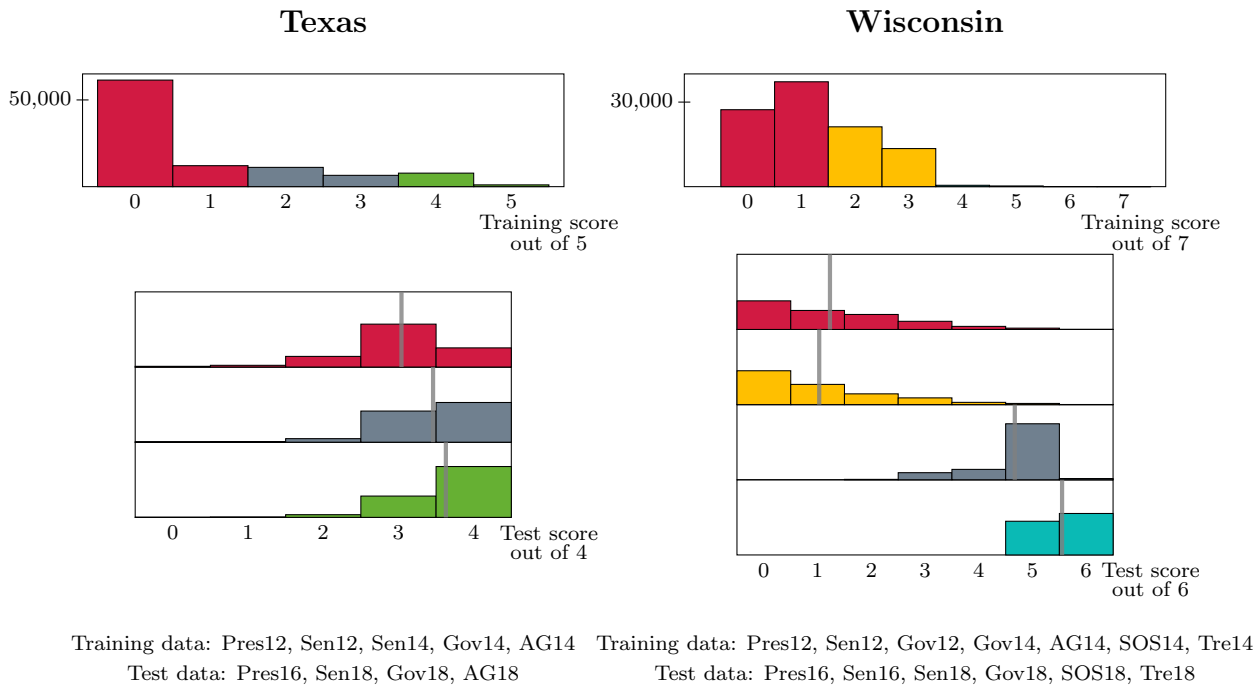
\centering
\begin{tikzpicture}
\begin{scope}[xshift=0,yshift=10.5cm]
\input{MD-stack}
\end{scope}
\begin{scope}[xshift=8cm,yshift=10.5cm]
\input{PA-stack}
\end{scope}
\begin{scope}[xshift=0,yshift=0cm]
\input{TX-stack}
\end{scope}
\begin{scope}[xshift=8cm,yshift=0cm]
\input{WI-stack}
\end{scope}
\end{tikzpicture}
\caption{Near-proportionality predictions, continued.  All four of these states show an overall tendency for near-proportionality ($t=.07$) on earlier elections to secure improved proportionality on later elections. With a minor exception in Wisconsin, where plans with a training score of 0-1 slightly outperform those that score 2-3, higher training scores always secure higher average test scores. }\label{fig:hists2}
\end{figure}

Given a plan that is shown to materially favor one party with respect to recent electoral data under a proportionality standard, is that property likely to persist into the future?
Conversely, given a plan that secures proportional outcomes with respect to several past elections, how confident can we be that it will continue to do so? 
An empirical investigation of this question, starting again with North Carolina, is shown in Figures \ref{fig:hists1}-\ref{fig:hists2}.\footnote{Note that these plots use a larger dataset than the designated FTV elections, with Attorney General, Governor, Treasurer, Secretary of State, and Comptroller elections where those were available.  This allows us to get a richer set of observed vote patterns in a limited number of years. We are using a simpler standard of $t=.07$ here rather than the FTV standard of $t=\max(.07,1/k)$, but the results will be identical in PA and TX and for the others we have confirmed that the findings are similar either way (see \cite{PaperGithub} for fuller outputs).  Observed statewide elections are useful for stress-testing how a districting plan responds under different realistic voting patterns.  Under this logic, it is not at all necessary that the elections employed for testing resemble the geography of recent Congressional voting; rather, we can watch what happens to a map as the ``sea level" of voting patterns rises and falls.  Congressional patterns are subject to change as incumbents shift, strong candidates come and go, and there are regional or national waves in voter sentiment.  Using a wide range of statewide elections gives us a view of the structural properties of a map, which we regard as more informative for a ten-year range of future voting than any single averaged or imputed voting pattern.  See for instance \cite{Mass-ELJ}, where we see that an election index  would obscure the fence-out effect for Massachusetts Republicans that was present for a large number of individual elections, studied serially.\label{note:MA}} 

In these plots, the top histogram shows the breakdown of the ensemble by {\em training score}, which is the number of near-proportional outcomes from the earlier elections.
Below that are histograms showing the {\em test scores} (near-proportionality in later elections) for the subset of plans with each training score, in turn.  
(The language of ``training" and ``testing" is borrowed from machine learning, where one collection of data is used to fit a model and a second collection of data is used to test the success of the model at making a prediction or classification.)

For instance, the North Carolina ensemble has 36,711 plans with a training score of zero, shown in the \textbf{\color{alizarin}{red}} bar.  Below that, the \textbf{\color{alizarin}{red}} histogram breaks those plans down by test score,  showing that the most common test score is also zero, while the gray line marks their average test score at 0.59.  There are 39 plans with a training score of 4, barely visible in the top histogram, and expanded in \textbf{\color{anothergreen}green} below (average test score 3.72).  Texas and Wisconsin had more training score values achieved than the other states,  so the training scores are grouped in pairs.

The results are quite encouraging.  In several states---North Carolina, Maryland, and Wisconsin---there is a pronounced trend for plans that are more often near-proportional in an earlier electoral window to stay so in a subsequent window.  In Texas and Pennsylvania, there is not much variation in the test score, but the small differences incline the right way.
Massachusetts, as usual, bucks the trend due to the uniform partisan geography in the later set of elections.  


\FloatBarrier
\section{Finding proportional alternatives}

In this section we present summary data showing that alternative plans can succeed from the point of view of average performance on the largest dataset available, as a complementary view to the serial performance considered above.  We choose to average {\em scores} rather than averaging the elections into a blended index. Election averaging, especially for a range of contests up and down the ballot, is fraught with issues of proper weighting given variable turnout, and can obscure consistent patterns that are important for understanding likely outcomes (see footnote \ref{note:MA} above).   

When we do compute a mean disproportionality,  the variance is also informative.
If the state dataset includes elections that are variable in their party lean, then the mean and the variance of \disprop combine to give a good view of a plan's properties.
Low-variance plans with $\disprop\approx 0$ across a diverse set of elections would be especially strong choices under a norm of proportionality.  By contrast, the most egregious---and most successful---gerrymanders would have high-magnitude $\disprop$ and low variance.  

\begin{figure}
    \centering
\begin{tikzpicture}
\node at (0,12) {\includegraphics[width=2.4in]{images/NC_scatter.png}};
\node at (-3.6,12) [rotate=90] {Variance};
\node at (-2,12+2.5) {\Large \bf NC};
\node at (7,12) {\includegraphics[width=2.4in]{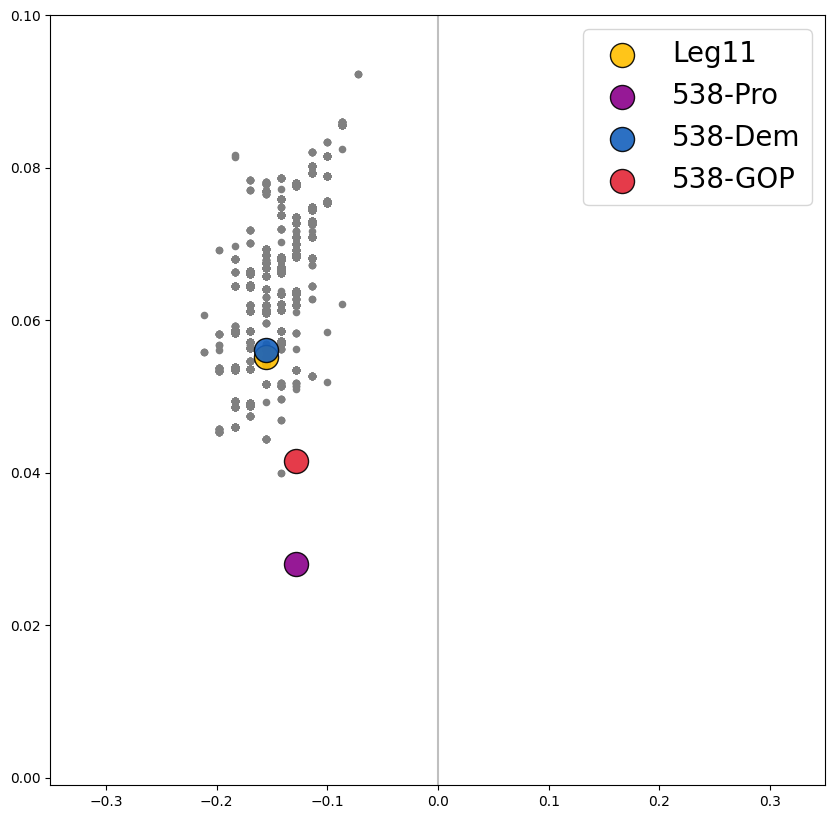}};
\node at (7-2,12+2.5) {\Large \bf MA};
\node at (0,6) {\includegraphics[width=2.4in]{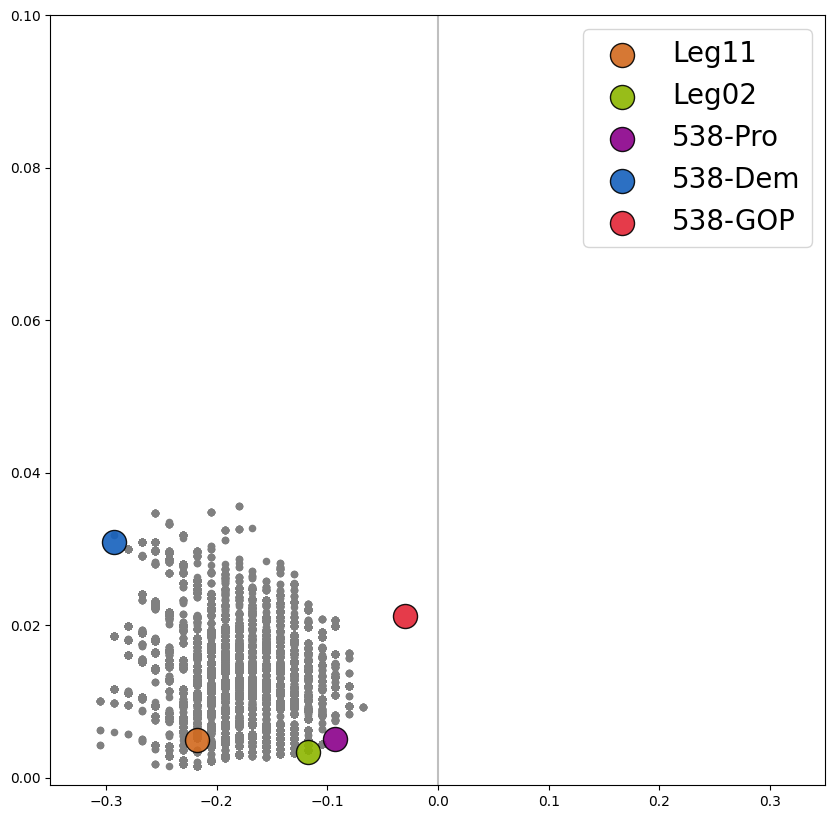}};
\node at (-3.6,6) [rotate=90] {Variance};
\node at (0-2,6+2.5) {\Large \bf MD};
\node at (7,6) {\includegraphics[width=2.4in]{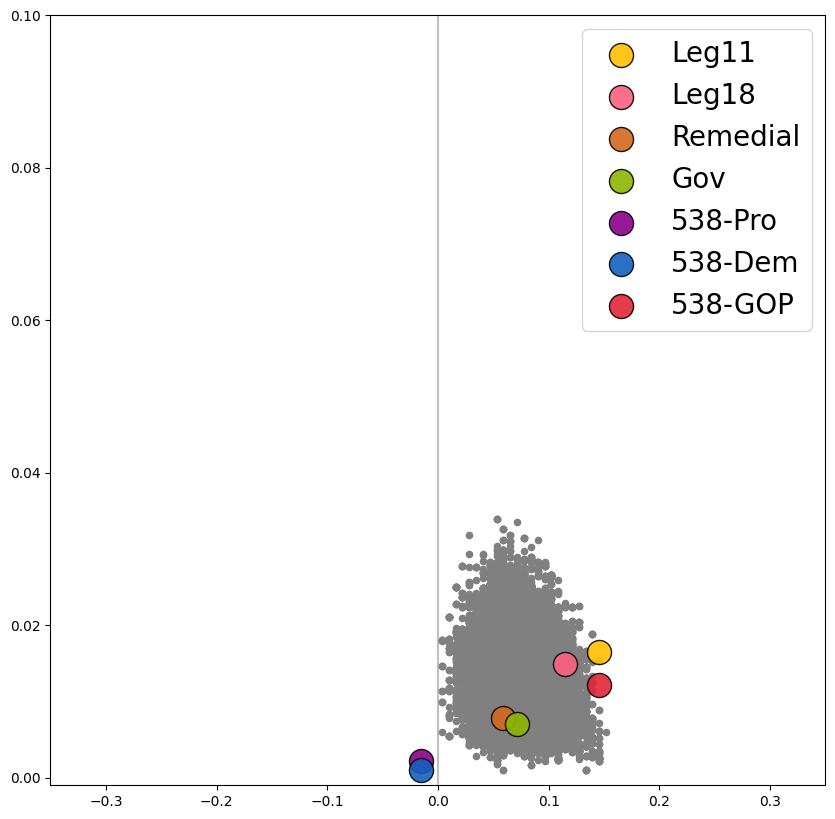}};
\node at (7-2,6+2.5) {\Large \bf PA};
\node at (0,0) {\includegraphics[width=2.4in]{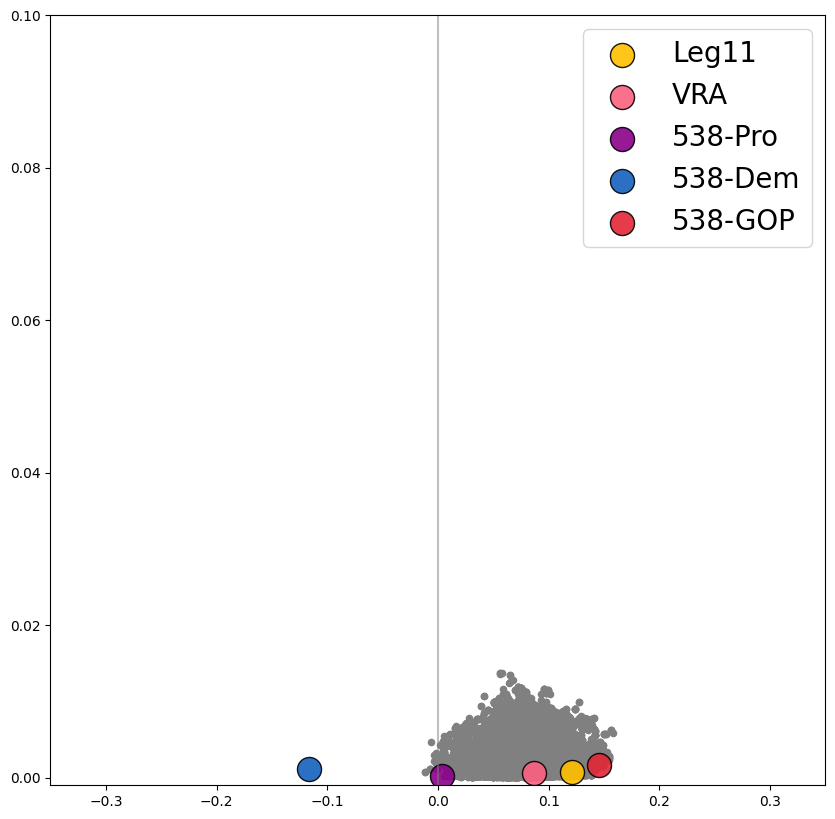}};
\node at (-2,2.5) {\Large \bf TX};
\node at (0,-3.4) {Mean \disprop};
\node at (-3.6,0) [rotate=90] {Variance};
\node at (7,0) {\includegraphics[width=2.4in]{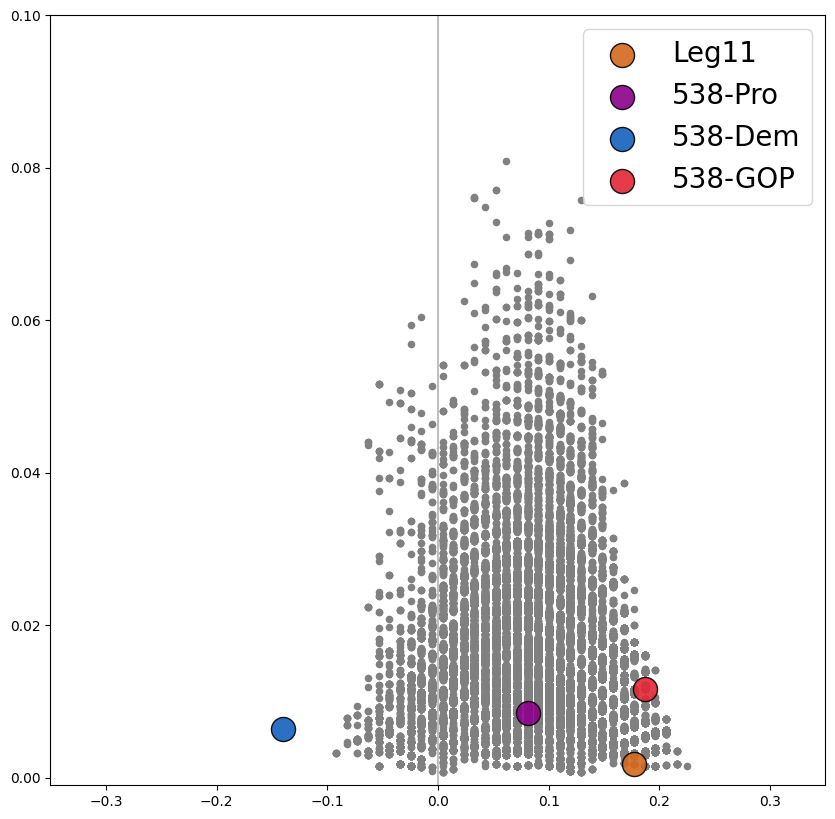}};
\node at (7-2,2.5) {\Large \bf WI};
\node at (7,-3.4) {Mean \disprop};
\end{tikzpicture}
    \caption{The gray scatterplots show the mean and variance of \disprop for each state's ensemble of 100,000 plans. (The MA dataset looks smaller because many of the points coincide.) The colored dots show the corresponding data for plans made by people, including legislators, courts, and journalists.  Besides legislative proposals and the 538 plans, these include a plan made by a bipartisan panel of retired judges in NC; a plan proposed by the Governor and a remedial plan ultimately adopted by the court in PA; and a limited modification of the legislative plan in TX that was drawn by a court (in the name of VRA compliance).  Colored dots are vertically jittered for visibility. NC plot is repeated from Figure~\ref{fig:NC-scatter}.}
    \label{fig:all-scatters}
\end{figure}

In Figure~\ref{fig:all-scatters}, we present the data corresponding to North Carolina's Figure~\ref{fig:NC-scatter} for other states under discussion.  
As in that figure, we include the Wasserman/538 Proportional, Democratic, and Republican plans.
Notably, the 538 project used a different method to measure proportionality, using an election index based on averaging Presidential contests to designate a target seat split; this is then paired with Congressional voting history to gauge the lean of the seats.\footnote{``This [proportionally partisan] map seeks to draw districts that favor each party in proportion to the overall political lean of each state. For example, if a state has 10 districts and Republicans won an average of 70 percent of its major-party votes in the last two presidential elections, we drew seven districts to favor Republicans and three to favor Democrats.... The probabilities of electing a Democrat or Republican are based on how often seats with a given Cook PVI elected members of each party between 2006 and 2016."}
Thus, if the 538 proportionally partisan (``538-Pro") map succeeds in FTV Test terms, it is a good sign that the FTV standard is robust to different ways of aiming at proportionality.

 In Massachusetts, the enacted plan---which performs the same as 538-Dem---is typical of the ensemble.  Maryland's enacted plan is fairly D-favoring, with moderately low variance.  By contrast, the benchmark plan from the previous cycle (though it would have become malapportioned in the intervening years) behaves similarly to the 538 Proportional map. In Pennsylvania, the 538 team used the same plan as both a Democratic and Proportional map. It is indeed fairly proportional, with $\disprop\approx 0$ and low variance.  The 2011 enacted plan, the legislators' proposed replacement plan from 2018, and the 538 Republican gerrymander are all strongly R-favoring, somewhat mitigated by moderate variance. The Texas 538 Proportional plan once again lives up to its name.  The court's VRA remedial plan displays a consistent Republican skew, with low variance, but the 538 Republican plan goes further.  
And Wisconsin follows the trend that a plan aimed at proportionality can succeed.

We can make a few other observations from the scatterplots.
For instance, the Massachusetts scatterplot shows considerably higher variance than in the other states because it includes not only President and Senate races but also others such as Governor's races, which can incline quite Republican in an otherwise Democratic-leaning state. 
The larger states (PA, TX) have more consistent statewide election shares, which contributes to lower variance in disproportionality.\footnote{It is also consistent with the findings of Rodden--Weighill in \cite{RoddenWeighill} that larger numbers of districts lead to a tighter range of observed seat shares.  That paper finds a remarkable stability in the partisan lean of ensembles as the scale varies, but with lower variance as the districts get more numerous.}


\FloatBarrier

\section{Conclusion}

The 50 United States are often called ``laboratories of democracy," and they are constitutionally entitled to construct state-specific legal frameworks for the ``times, places, and manner" of elections, within the bounds of federal law.  The {\em manner} includes a choice of systems (such as plurality elections in single-member districts, or PSMD) and a choice of norms for redistricting.  For Congress, 45 of 50 states conducted PSMD elections in 2022.\footnote{See Article 1 of the U.S. Constitution (\href{https://constitution.congress.gov/browse/article-1/section-4/clause-1/}{\tt constitution.congress.gov}).
Federal law currently requires the use of single-member districts for the U.S. House of Representatives, but does not specify plurality as an election mechanism.  Three states, California, Georgia, and Louisiana, have a runoff system in place for general elections if the plurality winner did not secure a majority. Two states, Maine and Alaska, now use a form of ranked choice voting in place of plurality in their districts.  In November 2022, Nevada voters approved Question 3, which would bring a third state to the ranked choice list if voters approve it again in 2024.}    In this paper, we explore whether it might be compatible with PSMD for states to adopt proportionality as a {\em goal} for Congressional districting.  We find evidence that the ease of securing a series of proportional outcomes varies by state, but that even party-blind redistricting can readily produce many examples in most states that we study---which includes all of the major legal hotspots from the previous Census cycle.  

Our empirical work clarifies how to perform the simple ``FTV Test" for extreme partisan gerrymanders that was set down in the draft legislation of the Freedom to Vote Act, and we confirm that the test is neither vacuous nor overly stringent.    Applied with a proportionality target, the test requires that at least 3 out of 4 designated elections produce near-proportional outcomes; if not, then a reason for deviation must be provided to a court.  
We consider five states that were widely regarded as partisan gerrymanders and challenged in court, and one more (Massachusetts) that was not but that has a notable political geography.
The results are summarized in Tables~\ref{tab:legplans} and \ref{tab:altplans}.\footnote{The four FTV contests are selected as though the test were being conducted at the end of 2018, when litigation was active in five of these states.  They are 
represented in the table in the following order:
NC (Pres12-Sen14-Pres16-Sen16); MD (Pres12-Pres16-Sen16-Sen18); PA (Pres12-Sen12-Pres16-Sen16); TX (Pres12-Sen14-Pres16-Sen18); WI (Pres12-Pres16-Sen16-Sen18). Ensemble25 reports the success of the 25th percentile plan, and Ensemble10 the 10th percentile:  for example, at least 10\% of ensemble plans pass (so Ensemble10 earns a P) in every state but MA, and at least 25\% of plans pass (so Ensemble25 earns a P) in every state but MA and PA.  
The pattern of $+/-/\checkmark$ notations in the Ensemble rows shows the most common configuration observed; for instance, the notation $\checkmark \checkmark + \checkmark$ in the Wisconsin/Ensemble25 row indicates that among Wisconsin plans with three near-proportional elections, the one most commonly failed is Sen16, missing in a Republican-favoring direction.}

\begin{table}[htb!]\centering
\begin{tabular}{|cc|cccc|c|}
\hline 
state & control & \multicolumn{4}{|c|}{FTV elections} & P/F \\
\hline \hline 
NC 2012& R & $+$ & $+$ & $+$ & $+$ & F\\
NC 2016& R & $+$ & $+$ & $+$ & $+$ & F\\
\hline  \hline 
MA & D & $-$ & $-$ & $-$ & $-$ & F\\
MD & D & $-$ & $-$ & $-$ & $-$ & F \\
PA 2011& R & $+$ & $\checkmark$ & $+$ & $+$ & F \\
PA 2018& R & $+$ & $+$ & $+$ & $+$ & F \\
TX & R & $+$ & $+$ & $+$ & $+$ & F \\
WI & R & $+$ & $+$ & $+$ & $+$ & F \\
\hline 
\end{tabular}\caption{Plans made by legislatures.   The $\checkmark$ signs indicate whether the disproportionality is below threshold; if not, the $\pm$ signs tell us which side is favored by the skew.  None of these legislative plans passes the FTV test overall by succeeding in at least three of the four designated elections.
\label{tab:legplans}}
\end{table}

\begin{table}[ht]\centering 
\begin{tabular}{|cc|cccc|cc|}
\hline 
state & alternative & \multicolumn{4}{|c|}{FTV elections} & P/F & rebut?\\
\hline \hline 
\multirow{4}{*}{NC} & Judges & $+$ & $+$ & $+$ & $+$ & F& \multirow{4}{*}{no}\\
 & 538-Pro & $\checkmark$ & $\checkmark$ & $\checkmark$& $\checkmark$ & P&   \\
 & Ensemble25 & $\checkmark$ & $\checkmark$ & $\checkmark$& $+$ & P& \\
 & Ensemble10 & $\checkmark$& $\checkmark$ & $\checkmark$ &$\checkmark$ & P& \\
\hline  \hline 
\multirow{3}{*}{MA} & 538-Pro & $-$ & $-$ & $-$ & $-$ & F& \multirow{3}{*}{strong} \\
 & Ensemble25 & $-$ & $-$ & $-$ & $-$ & F& \\
 & Ensemble10 & $-$ & $-$ & $-$ & $-$ & F& \\
\hline 
\multirow{4}{*}{MD}& 2002 & $\checkmark$ & $\checkmark$ &
 $\checkmark$ & $\checkmark$& P & \multirow{4}{*}{no}\\
 & 538-Pro & $\checkmark$ & $\checkmark$ & $\checkmark$& $\checkmark$ & P& \\ 
 & Ensemble25 & $\checkmark$& $\checkmark$ & $\checkmark$ &$-$ & P& \\
 & Ensemble10 & $\checkmark$& $\checkmark$ & $\checkmark$ &$-$ & P& \\
\hline 
\multirow{5}{*}{PA} & Remedial & $\checkmark$ & $+$ & $\checkmark$ &$\checkmark$ & P & \multirow{5}{*}{weak} \\
 & 538-Pro & $\checkmark$ & $\checkmark$ & $\checkmark$& $\checkmark$ & P& \\
 & Gov (D) & $+$ & $\checkmark$ & $+$ &$+$ & F &  \\
 & Ensemble25 & $\checkmark$& $\checkmark$ & $+$ &$+$ & F& \\
 & Ensemble10 & $\checkmark$& $\checkmark$ & $\checkmark$ &$+$ & P& \\
\hline 
\multirow{4}{*}{TX} & VRA & $\checkmark$ & $\checkmark$ & $+$ & $+$ & F & \multirow{4}{*}{no}\\
 & 538-Pro & $\checkmark$ & $\checkmark$ & $\checkmark$ & $\checkmark$ & P & \\
 & Ensemble25 & $\checkmark$& $+$ & $\checkmark$ &$\checkmark$ & P& \\
& Ensemble10 & $\checkmark$& $+$ & $\checkmark$ &$\checkmark$ & P& \\
\hline 
\multirow{3}{*}{WI}  & 538-Pro & $\checkmark$& $\checkmark$ & $\checkmark$& $\checkmark$ &  P & \multirow{3}{*}{no}\\
 & Ensemble25 & $\checkmark$& $\checkmark$ & $+$ &$\checkmark$ & P& \\
 & Ensemble10 & $\checkmark$& $\checkmark$ & $\checkmark$ &$\checkmark$ & P& \\
\hline 
\end{tabular}
\caption{Here, we consider alternative plans that are plausibly more proportional, though they were not necessarily made with that goal in mind (particularly the Judges' plan in NC, the Governor's  plan in PA, and the VRA plan in TX). 
For the ensemble plans, we look at the 25th percentile and 10th percentile performance, as explained in the text.
The last column is a note on whether this evidence might help the state rebut the presumption of a gerrymander.  Massachusetts would have a very strong case for rebuttal. 
Pennsylvania might attempt a rebuttal of the presumption of gerrymandering, but the evidence clearly favors the finding that their plans are impermissible under this standard.
We see no reasonable grounds to attempt rebuttal in the other states. \label{tab:altplans}}
\end{table}

In Massachusetts, the legislature's plan scores 0 out of 4 with all deviations in a Democratic direction, but in fact every one of our 100,000 blind plans scores 0 out of 4 in the same way, indicating that this disproportionality is explained (and likely forced) by  political geography.  This provides ample evidence to rebut the presumption of gerrymandering, and it fits with earlier research findings in \cite{Mass-ELJ}.  The case of Massachusetts illustrates the utility of setting up the FTV test as a rebuttable presumption of gerrymandering, so as to avoid requiring what is potentially impossible.   
In the other five states, the 538 Proportional plan and at least 1/10 of a neutral districting ensemble pass the test.  As noted above, the randomized algorithms employed here explore the central tendencies of the PSMD system, and it is common for a mapmaker with a goal to produce plans that surpass the extremes of a neutral ensemble.  Together, the 538 and ensemble evidence should give a court ample grounds to deny an attempted rebuttal in these five states.

\FloatBarrier

Proportionality is a straightforward and intuitive normative standard, and in most states studied here, it is achievable in a single-member/plurality district system without extreme partisan tuning---in every state but Massachusetts, a substantial share of maps from a partisan-blind process pass the test.
Taken together, our findings provide ample evidence that the FTV Test for redistricting is sound when used with a proportionality target (and a safety valve of rebuttability), in that it is both manageable and reasonably comports with the ``ground truth" determinations from courts that identified certain plans as impermissible partisan gerrymanders.  This should provide encouraging empirical support for 
 policymakers to reintroduce a similar test at the federal level, or  to specify a proportionality target at the state level.

Though proportionality is not to be automatically expected from plurality in single-member districts, even absent a gerrymandering agenda,  the findings here should raise our confidence that in many cases proportionality can nonetheless be  pursued---and achieved.

\newpage
\appendix

\section{Ensemble specs}\label{app:recom}

This paper uses ensembles of 100,000 districting plans produced by the {\em recombination} (``ReCom") Markov chain method described in \cite{ReCom} and implemented by the MGGG Redistricting Lab in Python \cite{GerryChain}.  Chains were run with uniform spanning trees and cut-edge district selection.  This approximately targets the {\em spanning tree distribution} on plans, which has the property that the ratio of the probabilistic weight given to two plans only depends on shape statistics (in this case, {\em spanning-tree compactness}, or the product of the number of spanning trees over the districts) and not on any other features of the particular geography and population distribution.  

The ensembles used here were made without a priority on county or municipality intactness, and with each district allowed to deviate no more than 1\% from ideal population.  Contiguity was enforced at every step.  Ensemble scripts, a variety of outputs and plots, and other replication materials may be found in the GitHub repository associated to this paper \cite{PaperGithub}.



\section{Text from the Freedom to Vote Act}\label{app:FTV}

Here is the text in full from \S 5003(c), the section dealing with partisan fairness, in S.2747 (Freedom to Vote Act) \cite{FTV}.

\begin{quote}
{\tt 
(c) No Favoring Or Disfavoring Of Political Parties.{\rm ---}

(1) PROHIBITION.{\rm ---} A State may not use a redistricting plan to conduct an election that, when considered on a statewide basis, has been drawn with the intent or has the effect of materially favoring or disfavoring any political party.

(2) DETERMINATION OF EFFECT.{\rm ---}The determination of whether a redistricting plan has the effect of materially favoring or disfavoring a political party shall be based on an evaluation of the totality of circumstances which, at a minimum, shall involve consideration of each of the following factors:

(A) Computer modeling based on relevant statewide general elections for Federal office held over the 8 years preceding the adoption of the redistricting plan setting forth the probable electoral outcomes for the plan under a range of reasonably foreseeable conditions.

(B) An analysis of whether the redistricting plan is statistically likely to result in partisan advantage or disadvantage on a statewide basis, the degree of any such advantage or disadvantage, and whether such advantage or disadvantage is likely to be present under a range of reasonably foreseeable electoral conditions.

(C) A comparison of the modeled electoral outcomes for the redistricting plan to the modeled electoral outcomes for alternative plans that demonstrably comply with the requirements of paragraphs (1), (2), and (3) of subsection (b) in order to determine whether reasonable alternatives exist that would result in materially lower levels of partisan advantage or disadvantage on a statewide basis. For purposes of this subparagraph, alternative plans considered may include both actual plans proposed during the redistricting process and other plans prepared for purposes of comparison.

(D) Any other relevant information, including how broad support for the redistricting plan was among members of the entity responsible for developing and adopting the plan and whether the processes leading to the development and adoption of the plan were transparent and equally open to all members of the entity and to the public.

(3) REBUTTABLE PRESUMPTION.{\rm ---}

(A) TRIGGER.{\rm ---}In any civil action brought under section 5006 in which a party asserts a claim that a State has enacted a redistricting plan which is in violation of this subsection, a party may file a motion not later than 30 days after the enactment of the plan (or, in the case of a plan enacted before the effective date of this Act, not later than 30 days after the effective date of this Act) requesting that the court determine whether a presumption of such a violation exists. If such a motion is timely filed, the court shall hold a hearing not later than 15 days after the date the motion is filed to assess whether a presumption of such a violation exists.

(B) ASSESSMENT.{\rm ---}To conduct the assessment required under subparagraph (A), the court shall do the following:

(i) Determine the number of congressional districts under the plan that would have been carried by each political party’s candidates for the office of President and the office of Senator in the 2 most recent general elections for the office of President and the 2 most recent general elections for the office of Senator (other than special general elections) immediately preceding the enactment of the plan, except that if a State conducts a primary election for the office of Senator which is open to candidates of all political parties, the primary election shall be used instead of the general election and the number of districts carried by a party’s candidates for the office of Senator shall be determined on the basis of the combined vote share of all candidates in the election who are affiliated with such party.

(ii) Determine, for each of the 4 elections assessed under clause (i), whether the number of districts that would have been carried by any party’s candidate as determined under clause (i) results in partisan advantage or disadvantage in excess of 7 percent or one congressional district, whichever is greater, as determined by standard quantitative measures of partisan fairness that relate a party's share of the statewide vote to that party's share of seats.

(C) PRESUMPTION OF VIOLATION.{\rm ---}A plan is presumed to violate paragraph (1) if it exceeds the threshold described in clause (ii) of subparagraph (B) with respect to 2 or more of the 4 elections assessed under such subparagraph.

(D) STAY OF USE OF PLAN.{\rm ---}Notwithstanding any other provision of this title, in any action under this paragraph, the following rules shall apply:

(i) Upon filing of a motion under subparagraph (A), a State’s use of the plan which is the subject of the motion shall be automatically stayed pending resolution of such motion.

(ii) If after considering the motion, the court rules that the plan is presumed under subparagraph (C) to violate paragraph (1), a State may not use such plan until and unless the court which is carrying out the determination of the effect of the plan under paragraph (2) determines that, notwithstanding the presumptive violation, the plan does not violate paragraph (1).

(E) NO EFFECT ON OTHER ASSESSMENTS.{\rm ---}The absence of a presumption of a violation with respect to a redistricting plan as determined under this paragraph shall not affect the determination of the effect of the plan under paragraph (2).

(4) DETERMINATION OF INTENT.{\rm ---}A court may rely on all available evidence when determining whether a redistricting plan was drawn with the intent to materially favor or disfavor a political party, including evidence of the partisan effects of a plan, the degree of support the plan received from members of the entity responsible for developing and adopting the plan, and whether the processes leading to development and adoption of the plan were transparent and equally open to all members of the entity and to the public.

(5) NO VIOLATION BASED ON CERTAIN CRITERIA.{\rm ---}No redistricting plan shall be found to be in violation of paragraph (1) because of the proper application of the criteria set forth in paragraphs (1), (2), or (3) of subsection (b), unless one or more alternative plans could have complied with such paragraphs without having the effect of materially favoring or disfavoring a political party.
}
\end{quote}

\clearpage
\section{FTV Test with other standards}


\begin{figure}[htb!]\centering 
\begin{tikzpicture}[scale=.395]
\begin{scope}[yshift=15cm]
\draw [ultra thick] (-6,-4) rectangle (16+6,7+4);
\node at (-7.4,3.5) [rotate=90] {\large \bf Proportionality};
\pie[hide number,pos={0,7},color={anothergreen,anotherblue,slategray,amber,alizarin}]{12.0/,13.9/,22.3/,18.4/,33.4/} 
\node at (0,7) [left=1.2cm,above=1cm]  {\bf NC};
\pie[hide number,pos={8,7},color={alizarin}]{100.0/}
\node at (8,7) [left=1.2cm,above=1cm]  {\bf MA};
\pie[hide number,pos={16,7},color={anothergreen,anotherblue,slategray,amber,alizarin}]{7.3/,54.2/,11.8/,3.8/,23.0/}
\node at (16,7) [left=1.2cm,above=1cm] {\bf MD};
\pie[hide number,pos={0,0},color={anothergreen,anotherblue,slategray,amber,alizarin}]
{0.6/,12.5/,34.6/,44.2/,8.1/}
\node at (0,0) [left=1.2cm,above=1cm]  {\bf PA};
\pie[hide number,pos={8,0},color={anothergreen,anotherblue,slategray,amber,alizarin}]
{1.4/,26.6/,59.5/,11.0/,1.5/}
\node at (8,0) [left=1.2cm,above=1cm]  {\bf TX};
\pie[hide number,pos={16,0},color={anothergreen,anotherblue,slategray,amber,alizarin}]
{20.2/,22.9/,40.1/,15.0/,1.8/}
\node at (16,0) [left=1.2cm,above=1cm]  {\bf WI};
\end{scope}

\begin{scope}
\draw [ultra thick] (-6,-4) rectangle (16+6,7+4);
\node at (-7.4,3.5) [rotate=90] {\large \bf Efficiency gap};
\pie[hide number,pos={0,7},color={anothergreen,anotherblue,slategray,amber,alizarin}]{27.0/,19.8/,21.7/,15.2/,16.3/}


\pie[hide number,pos={8,7},color={anothergreen,anotherblue,slategray,amber,alizarin}]{0.0/,0.0/,0.0/,0.0/,100.0/}
\pie[hide number,pos={16,7},color={anothergreen,anotherblue,slategray,amber,alizarin}]{52.4/,33.7/,3.8/,5.0/,5.0/}

\pie[hide number,pos={0,0},color={anothergreen,anotherblue,slategray,amber,alizarin}]{0.3/,9.6/,26.9/,33.9/,29.3/}

\pie[hide number,pos={8,0},color={anothergreen,anotherblue,slategray,amber,alizarin}]{83.4/,14.9/,1.6/,0.1/,0.0/}

\pie[hide number,pos={16,0},color={anothergreen,anotherblue,slategray,amber,alizarin}]{20.2/,22.9/,40.1/,15.0/,1.8/}

\end{scope}

\begin{scope}[yshift=-15cm]
\draw [ultra thick] (-6,-4) rectangle (16+6,7+4);
\node at (-7.4,3.5) [rotate=90] {\large \bf Ensemble mean};

\pie[hide number,pos={0,7},color={anothergreen,anotherblue,slategray,amber,alizarin}]{47.6/,27.0/,13.7/,7.6/,4.1/}


\pie[hide number,pos={8,7},color={anothergreen,anotherblue,slategray,amber,alizarin}]{100.0/,0.0/,0.0/,0.0/,0.0/}


\pie[hide number,text=legend,pos={16,7},color={anothergreen,anotherblue,slategray,amber,alizarin}]{78.4/4,10.6/3,3.9/2,7.2/1,0.0/0}


\pie[hide number,pos={0,0},color={anothergreen,anotherblue,slategray,amber,alizarin}]{52.5/,36.7/,9.6/,1.2/,0.0/}


\pie[hide number,pos={8,0},color={anothergreen,anotherblue,slategray,amber,alizarin}]{86.1/,12.4/,1.5/,0.1/,0.0/}


\pie[hide number,pos={16,0},color={anothergreen,anotherblue,slategray,amber,alizarin}]{57.1/,34.2/,7.4/,1.2/,0.1/}
\end{scope}

\end{tikzpicture}
\caption{How easy is it to pass the FTV Test with different partisan fairness metrics setting the target?  We compare the shares of a 100,000-map ensemble that are close to ideal with respect to proportionality (repeated from Figure~\ref{fig:pie} above), efficiency gap, and the ensemble mean, respectively, recalling that the \textbf{\color{anothergreen}{green}} and \textbf{\color{anotherblue}{blue}} sectors pass the test.  An $EG$ target is more commonly hit than a proportionality target in NC, MD, and especially TX, while there is no difference in MA and WI, and near-proportionality is slightly more common in PA. (In PA, but not in WI, one of the four test elections has far enough from even vote share that the two standards diverge.)
Finally, the large \textbf{\color{anothergreen}{green}} sectors in the bottom set of charts reflect the expected finding that a major share of an ensemble falls close to the ensemble mean.}\label{fig:three-sets-of-pies}
\end{figure}
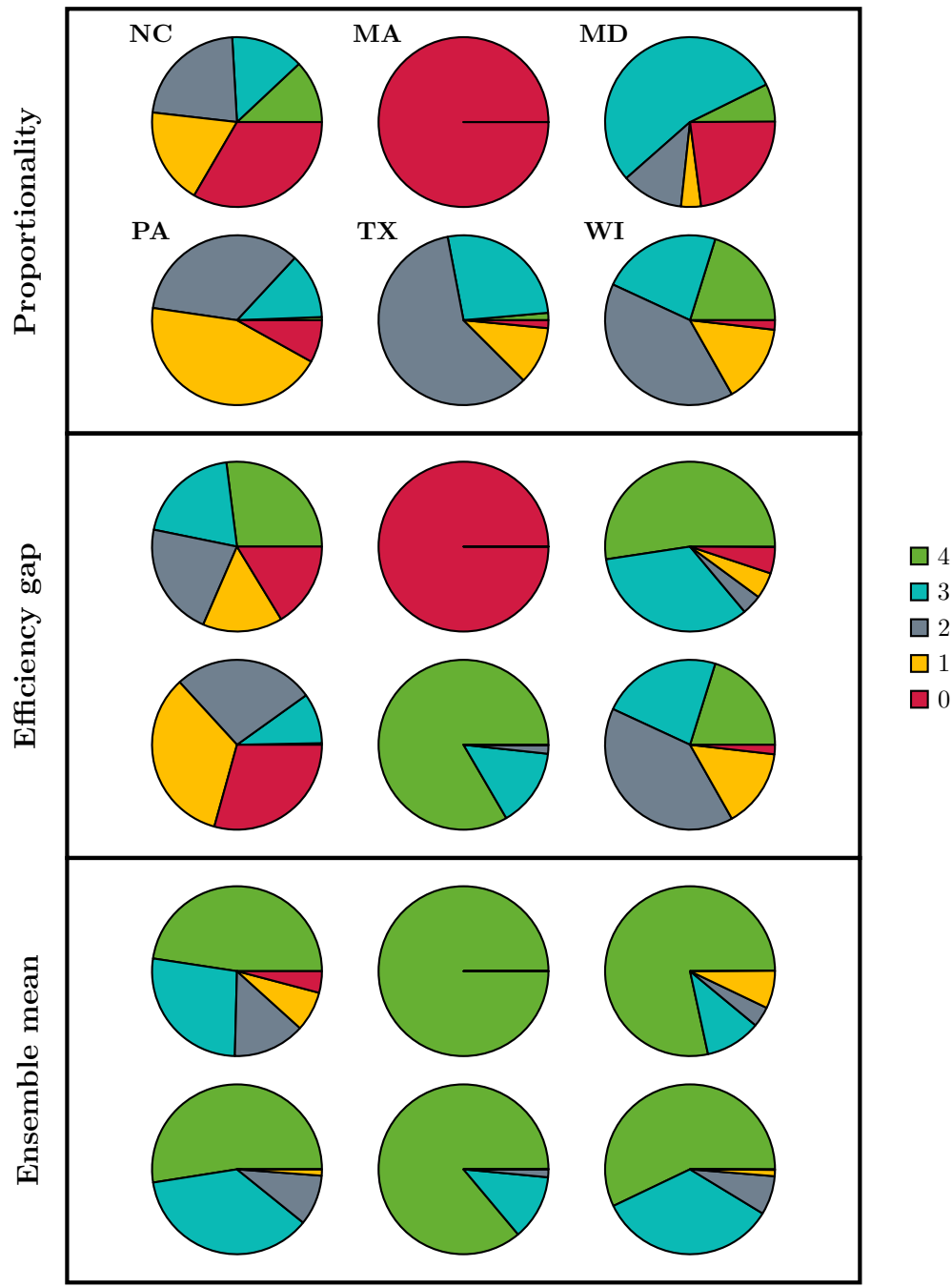

\end{document}